\documentclass{ws-procs9x6}

\begin{document}

\title{THE NUCLEON ELECTRIC DIPOLE MOMENT IN LIGHT-FRONT QCD}

\author{S. Gardner}

\address{Department of Physics and Astronomy, University of Kentucky,\\
Lexington, KY, USA \\
E-mail: gardner@pa.uky.edu }

\begin{abstract}
I present an exact relationship between
the electric dipole moment and anomalous magnetic moment 
of the nucleon in the light-front formalism of QCD
and consider its consequences. 
\end{abstract}

\bodymatter

\section{Introduction}
Interpreting the electric dipole moments of leptons and baryons 
as constraints on fundamental, CP-violating Lagrangian parameters of
various extensions of the Standard Model (SM) gives key insight
into TeV-scale physics. In this contribution I report 
on work in collaboration with Stan Brodsky and 
Dae Sung Hwang~\cite{bghedm}, in
which we sharpen the connection between the computed 
neutron electric dipole moment and fundamental CP violation by
comparing its hadronic matrix element to that of the anomalous
magnetic moment. 
In the context of the SM, and, specifically, of
the Cabbibo-Kobasyashi-Maskawa (CKM) mechanism of CP violation, 
the assessed values of the neutron electric dipole moment (EDM) have 
been disparate, ranging from 
$d^n_{\rm CKM} \simeq 10^{-32}\hbox{e-cm}$~\cite{gavela,KZ82} 
arising from a $\pi-N$ loop calculation in 
a chiral Lagrangian treatment, to 
$d^d_{\rm KM} \simeq 10^{-34}\hbox{e-cm}$~\cite{krip,czar} for the 
EDM of the $d$-quark itself, computed to three-loop precision in 
leading-logarithmic approximation. These EDMs are much too small 
to be experimentally observable, so that the marked disparity is 
actually of little
consequence. However, if we restrict ourselves to effective 
CP-violating operators of dimension five or less, the 
method of QCD sum rules can be employed 
to compute the EDM of the neutron~\cite{popritz,edmrev05},  
yielding a value for $d^n$, induced by a QCD $\bar\theta$-term, e.g.,
commensurate in size with that of the chiral estimate~\cite{crewther}, 
with a surety of $\sim 50\%$~\cite{popritz}.  
The evaluation of $d^n$ and $d^p$, and the errors therein, 
is also important to interpreting the $^2H$ EDM~\cite{lebedev}. 
Here we analyze the nucleon electric dipole moment 
in the light-front formalism of QCD~\cite{bghedm}, to the end of 
realizing an independent test of 
the methods used to compute $d^N$ --- and of their assessed errors. 

\section{Electromagnetic Form Factors in Light-Front QCD}
Our study of the electric dipole form factor $F_3(q^2)$
in the light-front formalism of QCD complements earlier
studies of the Dirac and Pauli form factors~\cite{lcff}. 
The Pauli and electric dipole form factors emerge from the 
spin-flip matrix elements 
of the electromagnetic current $J^\mu(0)$:
\begin{eqnarray}
\langle P', -S_z | J^\mu (0) |P, S_z\rangle
    &=& 
\bar U(P',-\lambda)\,\Bigg[\, {\frac{i}{2M}} \sigma^{\mu\alpha} 
\nonumber \\
&& 
\times 
\left( F_2(q^2)  + i F_3(q^2) \gamma_5 \right)q_\alpha\, 
\Bigg] \, U(P,\lambda)\,,
\end{eqnarray}
where $U(P,\lambda)$ is a Dirac spinor for a nucleon of momentum $P$
and helicity $\lambda$, with $S_z=\lambda/2$. 
Recall that the anomalous magnetic
moment $\kappa$ and the electric dipole moment $d$ are given by
$\kappa =(e/{2 M})[F_2(0)]$ and 
$d=({e}/{M})[F_3(0)]$. 
We find a close connection between $\kappa$ and $d$~\cite{bghedm},
as long anticipated~\cite{bigiural}. Working in the $q^+=0$ frame, 
with 
$q = (q^+,q^-,\mathbf{q}_{\perp}) = (0, -q^2/P^+,\mathbf{q}_{\perp})$
and
$P = (P^+,P^-,\mathbf{P}_{\perp}) = (P^+, M^2/P^+, \mathbf{0}_{\perp})$, 
in the interaction picture for $J^+(0)$, and in the assumed 
simple vacuum of the light-front formalism, we find, noting 
$q^{{R}/{L}} \equiv q^1 \pm i q^2$, 
\begin{eqnarray}
&&
\frac{1}{2M} \left( 
\begin{array}{c} F_2(q^2)\\ -i F_3(q^2) \end{array}
\right)
 =
\sum_a  \int
[{\mathrm d} x] [{\mathrm d}^2 \mathbf{k}_{\perp}]
\sum_j e_j \ \frac{1}{2}\ 
\Big[\,  
-\frac{1}{q^L}
\psi^{\uparrow *}_{a}(x_i,\mathbf{k}^\prime_{\perp
i},\lambda_i) \,
\nonumber \\
&& \times \psi^\downarrow_{a} (x_i, \mathbf{k}_{\perp i},\lambda_i) 
\pm  \frac{1}{q^R}
\psi^{\downarrow *}_{a}(x_i,\mathbf{k}^\prime_{\perp
i},\lambda_i) \,
\psi^\uparrow_{a} (x_i, \mathbf{k}_{\perp i},\lambda_i)\, \Big]
{}\, , 
\end{eqnarray}
where 
$\mathbf{k}'_{\perp j}=\mathbf{k}_{\perp j}+(1-x_j)
\mathbf{q}_{\perp}$ 
for the struck constituent $j$ and
$\mathbf{k}'_{\perp i}=\mathbf{k}_{\perp i}-x_i \mathbf{q}_{\perp}$
for each spectator ($i\ne j$). The electric dipole form factor $F_3(q^2)$
vanishes if the usual light-front wave functions are employed, so
that we must learn how parity- and time-reversal-violating
effects can be included in the light-cone framework. 

\section{Discrete Symmetries on the Light Front and a Relation for
the Electric Dipole Moment}
We construct parity ${\cal P}_\perp$
and time-reversal ${\cal T}_\perp$ in the light-front 
formalism~\cite{bghedm} by noting that these operations should act on
the $k_\perp$ of a free particle alone, so that 
$|\mathbf{k}_\perp|$, $k^+$, and $k^-$ remain unchanged. 
We choose ${\cal P}_\perp$ so that the components
of a vector $d^\mu$ transform as $d^1 \to - d^1$, 
$d^2 \to  d^2$, or $d^{R,L} \to - d^{L,R}$, 
and $d^\pm\to d^\pm$. With this 
${\cal P}_\perp$ is an unitary operator, though it flips the 
spin as well. We find  that 
$F_2(q^2)$ is even and $F_3(q^2)$ is odd under ${\cal P}_\perp$. 
The choice of ${\cal T}_\perp$ is predicated by that for 
${\cal P}_\perp$; a momentum vector $q^\mu$ transforms
as $q^{R,L} \to - q^{L,R}$ and $q^\pm\to q^\pm$ under 
${\cal T}_\perp$, so that the position vector 
$x^\mu \equiv (x^+, x^-, x^L, x^R) \to (-x^+, -x^-, x^R, -x^L)$. 
With this we find that ${\cal T}_\perp$
is antiunitary, but it does not flip the spin. With 
the charge-conjugation operator ${\cal C}$ 
defined in the usual way, we note that
all scalar fermion bilinears are invariant under 
${\cal C}{\cal P}_\perp {\cal T}_\perp$ as needed. 
We find that $\hbox{Re}(F_2)$ and $\hbox{Im}(F_3)$
are even
and $\hbox{Re}(F_3)$ and $\hbox{Im}(F_2)$
are odd under ${\cal T}_\perp$. 
Thus to realize a non-zero
electric dipole form factor, we must include a 
${\cal T}_\perp$- and ${\cal P}_\perp$-odd parameter $\beta_a$ 
in the light-front
wave function 
$\psi^{S_z}_{a} (x_i, \mathbf{k}_{\perp i},\lambda_i)$; namely, 
$\psi^{S_z}_{a} (x_i, \mathbf{k}_{\perp i},\lambda_i)
= \phi^{S_z}_{a} (x_i, \mathbf{k}_{\perp i},\lambda_i) 
\exp (i\lambda \beta_a)$, where 
$\phi^{S_z}_{a} (x_i, \mathbf{k}_{\perp i},\lambda_i)$ is both
${\cal P}_\perp$- and ${\cal T}_\perp$- invariant. 
We assume ${\cal C}{\cal P}_\perp$ 
is broken at scales much larger than those of 
interest, so that $M_{\rm CP}^2 \gg q^2$ and that any $q^2$-dependence
in $\beta_a$ can be neglected. With this we find, 
for a Fock component $a$~\cite{bghedm}, 
\begin{equation}
[F_3(q^2)]_a = (\tan\beta_a) [F_2 (q^2)]_a  
\quad \hbox{and} 
\quad d_a = 2\kappa_a  \beta_a
\quad \hbox{as} \quad q^2\to 0 \,,
\label{rel}
\end{equation}
since $\beta_a$ is small. Thus the EDM and the anomalous magnetic moment
of the nucleon should both be computed 
with a given method, to test for consistency. 
If the method employed is unable to confront the empirical anomalous
magnetic moments successfully, it cannot be trusted to predict the
electric dipole moments reliably. We note in the case of
the QCD sum rule method, that the computed 
anomalous magnetic moments, as long
known, are in good agreement with experiment, 
namely, $\kappa^n_{\rm th} = -2$ and $\kappa^p_{\rm th} = +2$~\cite{ioffe}.

\section{Implications}
We now consider some specific consequences of Eq.~(\ref{rel}). In what follows
we consider the EDM induced through a QCD $\bar\theta$-term only. 
In a quark--scalar-diquark, $q(qq)_0$, model of the nucleon a single 
${\cal P}_\perp$- and ${\cal T}_\perp$-violating parameter $\beta^N$ suffices
to characterize $d^N$. 
Since $\delta L_{CP}$ is isoscalar, $\beta^n = \beta^p$, and we
can employ the empirical anomalous magnetic moments 
$\kappa^n = -1.91$ and $\kappa^p = 1.79$, in units of $\mu_N$,
to estimate 
$(d^n + d^p)/(d^p - d^n) =
(\kappa^n + \kappa^p)/(\kappa^p - \kappa^n) \approx -0.12 /3.70\approx -0.03$. 
The isoscalar electric dipole moment of the nucleon is extremely small. This is
in accord with the chiral Lagrangian estimate, 
for which it is zero --- the relevant diagrams are 
mediated by a $\pi-N$ loop,
which is logarithmically enhanced as $M_\pi\to 0$~\cite{crewther}.  
Our estimate can be compared to the QCD sum rule calculation, for which 
$(d^n + d^p)/(d^p - d^n) \approx -0.3$~\cite{lebedev}, which is
much larger. We note that the QCD sum rule method also predicts a 
zero isoscalar magnetic moment~\cite{ioffe}; the method is less
successful in reproducing a quantity which suffers partial 
cancellation. The $^2 H$ magnetic moment is 
determined not only by the sum of $d^n + d^p$ but also by a 
CP-violating meson-exchange current --- 
the former is estimated to be numerically larger~\cite{lebedev}. 
The efficacy of
a EDM measurement in a particular system 
in bounding $\bar\theta$ is determined by
the size of the coefficient multiplying $\bar\theta$. 
The larger the coefficient, the better the bound on $\bar\theta$, 
for a given experimental limit.  Were $d^n+d^p$ smaller, 
the bound on $\bar\theta$ from a putative $^{2}H$ EDM measurement
would weaken. 

\section{Summary}
In summary, we have analyzed
the electromagnetic form factors
in the light-front formalism of QCD, extending the earlier
Drell-Yan-West-Brodsky framework to the analysis of
${\cal P}_\perp$ and ${\cal T}_\perp$-odd observables.
We have used the light-front formalism to find a
general equality between the anomalous magnetic and
electric dipole moments. The 
relation holds for spin-$1/2$ systems, in general: it is 
not specific to the 
neutron and is independent of the mechanism of CP violation.
An earlier study noting the
importance of the simultaneous study of the muon's electric dipole 
and anomalous magnetic moments is given in Ref.~\cite{feng}. 
The relation we derive implies that 
both the EDM and anomalous magnetic moment of the spin-$1/2$
system of interest should 
be calculated in a given model, to test for consistency. 
Ultimately, this can lead to sharpened constraints on 
models containing non-CKM sources of CP violation. 


\section*{Acknowledgments}
S.G. thanks S.J. Brodsky and D.-S. Hwang for 
a most enjoyable collaboration and acknowledges the support of
the U.S. Department of Energy under contract no. DE-FG02-96ER40989.



\bibliographystyle{aipproc}   

\end{document}